%% file: main.tex
\documentclass[submission,copyright,creativecommons]{eptcs}
\providecommand{\event}{SYNT 2015} 

\input{macros}

\newcommand\mytitle{Specification Format for Reactive Synthesis Problems}
\def\titlerunning{\mytitle}
\def\authorrunning{A. Khalimov}

\title{\mytitle\thanks{This work was supported by the Austrian Science Fund via project RiSE (S11406).}}

\author{Ayrat Khalimov\institute{Graz University of Technology, Austria}}

\begin{document}

\iffinal
\else
\li 
\- this is controller synthesis? \href{http://mbp.fbk.eu/scmbp.html} \href{http://mbp.fbk.eu/gpt2006sc.pdf}
\- enable using of assumptions F
\- format: allow defining the module to be synthesized (and partial information?)
\- allow the native SPEC -- check what smvflatten supports
\- check propositions used in automata are defined in smv file
\il
\fi

\maketitle

\begin{abstract}
Automatic synthesis from a given specification automatically constructs correct
implementation. This frees the user from the mundane implementation work, but still
requires the specification. But is specifying easier than implementing?  In
this paper, we propose a user-friendly format to ease the specification work, in
particularly, that of specifying partial implementations. Also, we provide
scripts to convert specifications in the new format into the SYNTCOMP format,
thus benefiting from state of the art synthesizers.
\end{abstract}


\input{intro}
\input{framework}

\input{examples}

\input{related}
\input{conclusions}
\bibliographystyle{eptcs}

\bibliography{crossrefs,refs}

\end{document}

%% file: macros.tex
\newif\iffinal
  \finaltrue

\usepackage{amsthm}
\usepackage{amsfonts}
\usepackage{amsmath}
\usepackage{amssymb}
\usepackage{temporal}
\usepackage{graphicx}
\usepackage{caption}
\usepackage{xspace}
\usepackage{subfig}

\usepackage{color,soul}
\definecolor{lightblue}{rgb}{.8,.95,1}

\usepackage{thmtools, thm-restate}

\usepackage{tikz}
\usepackage{pgf}
\usetikzlibrary{positioning,calc,automata,arrows,fit,shapes}

\usepackage{listings}

\let\note\relax

\iffinal
  \usepackage[finalold]{trackchanges} 
\else
  \usepackage[inline]{trackchanges}
\fi

\addeditor{$S\!J$}
\addeditor{$S\!A$}
\addeditor{$A\!K$}
\addeditor{$R\!B$}

\newcommand{\ak}[1]{\note[$A\!K$]{#1}}

\usepackage{hyperref}   

\newcommand{\mailto}[1]{\href{mailto:#1}{\nolinkurl{#1}}}

\definecolor{darkgreen}{rgb}{0,0.5,0}
\definecolor{darkblue}{rgb}{0,0,.5}
\definecolor{mygray}{gray}{.3}

\newcommand{\state}{q}

\newcommand{\stateset}{\expandafter\MakeUppercase\expandafter{\state}}
\newcommand{\State}{s}
\newcommand{\Stateset}{\expandafter\MakeUppercase\expandafter{\State}}

\newcommand{\impl}{\rightarrow}

\iffinal
\newcommand{\gray}[1]{}
\else
\newcommand{\gray}[1]{{\color{black!50} #1}}
\fi

\newcommand\G\always\xspace
\newcommand\F\eventually\xspace
\newcommand\W\weakuntil\xspace
\newcommand\U\until\xspace
\newcommand\X\nextt\xspace
\renewcommand{\GF}{\G\!\F\xspace}

\newcommand\li{\begin{itemize}}
\newcommand\il{\end{itemize}}
\renewcommand{\-}{\item}


%% file: intro.tex
\section{Introduction}            \label{sec:intro}

Specifying reactive synthesis tasks is not easy.
First, writing non-trivial specifications in e.g. linear temporal logic (LTL)
requires experience, and even an experienced user of LTL may notice that some
properties are easier to implement oneself than to specify.
Thus, it is desirable to be able to mix imperative and declarative paradigms
when specifying reactive synthesis tasks, which makes a call for a new
convenient specification format.

The full set of features of the new specification format might include:
\li
\-[1.] \emph{Modularity.}
A synthesis task may require to synthesize several communicating modules
where each module has its own properties.
Thus, the new format should allow for specifying module interfaces and 
connections between them.
These interfaces specify the amount of information each module knows about others.

\-[2.] \emph{Imperative and declarative.}
Some modules may already be given to the user, and some modules or parts of it
may be easier to implement than to specify.
Thus, the new format should allow for specifying module implementations.

\-[3.] \emph{Conversion to the SYNTCOMP format.}
The SYNTCOMP format~\cite{SYNTCOMP} was recently proposed as the common
ground format for reactive synthesis competitions, and at least four
synthesizers were competing in 2014.
Thus, to let the user to benefit from state of the art synthesizers, the new
format should be convertible into the SYNTCOMP format.

\-[4.] \emph{Property language agnostic.}
The new format should allow the user to choose the best suited language for
writing properties: linear temporal logic, linear dynamic logic~\cite{LDL},
regular expressions, automata, etc.
\il
These features requirements are our subjective suggestions and arise from the
domain of synthesis of reactive systems that usually represent some hardware.
The features certainly depend on the synthesis domain: for example, in the 
case of fault-tolerant algorithms the user also needs to specify the ratio of
faulty to normal processes, the type of faults, etc.

In this the paper we:
\li
\- propose a specification format for reactive synthesis tasks, and 
\- provide scripts to convert from the new format into the SYNTCOMP format.
\il

The new format can be extended to support features (1), (2), (3), and (4), 
but the current version has limitations. 
Some of the limitations are:
(i) the user can separate the system into modules, but each module has the full
information about others, (ii) only deterministic B\"uchi automata are allowed 
for specifying properties, and (iii) assumptions must be safety properties.

The new format is based on the SMV format~\cite{SMV} -- it is convenient for
describing hardware systems: it allows the user to define finite state machines that
operate on variables of enumeration and range types, and to separate the system
into modules, etc.
Another advantage of using the SMV format as the starting point is that there
is a solid support of the SMV format in the AIGER distribution~\cite{AIGER},
which greatly simplifies the task of the development of the conversion scripts.

\paragraph{Outline.}
We describe the new format and its restrictions in Section~\ref{sec:format}.
Section \ref{sec:scripts} describes the conversion scripts and also introduces
the SYNTCOMP format extended with liveness which is one of the supported target
formats (alongside the standard SYNTCOMP format).
Section~\ref{sec:example} illustrates the use of the format and of the scripts
-- we write the specification that describes the task: when given an
implementation of a Huffman decoder for the English alphabet, synthesize an
encoder for it.
Section~\ref{sec:related} points to other possible ways of writing specifications
and converting them into the SYNTCOMP format.
And we conclude in Section~\ref{sec:conclusions}.

%% file: framework.tex
\section{Specification Format}    \label{sec:format}

\ak{clarify: a memory-less strategy always exists(?)}

We assume that the reader is familiar with the SMV (cf.~\cite{SMV}) and the SYNTCOMP~\cite{SYNTCOMP} formats.
We introduce a new section into the SMV format, and the comments of special
form that allow for specifying synthesis problems.
The specification in the extended SMV format is then translated into the
SYNTCOMP format.

An example of the extended SMV format is shown in Listing~\ref{structure}.
\footnote{The format is under active development and may slightly differ from the one described here.} 

\begin{figure}[h]
\lstdefinelanguage{SMV}
{
keywords={module,var,define,syntspec},
sensitive=false,
keywordstyle=\color{orange},
morecomment=[l][\color{blue}]{--},
morecomment=[l][\color{gray}]{//},
morecomment=[s][\color{blue}]{SYS_AUTOMATON}{_SPEC},
morecomment=[s][\color{blue}]{ENV_AUTOMATON}{_SPEC},
}
\lstset{language=SMV}
\begin{lstlisting}[
basicstyle=\ttfamily\scriptsize,
frame=single, 
caption={Format structure (special elements are in {\color{blue} blue} color).},
label=structure]
MODULE helper1(input1,input2)  //we can define and use SMV modules as usually
VAR 
  state: 0..100;
DEFINE 
  reached42 := state=42;
  ...

MODULE main  // module `main' contains a specification
VAR  
  CPUread: boolean;   // only boolean is allowed

VAR --controllable
  valueOut: boolean;  // only boolean is allowed

VAR
  h: helper1(readA, valueOut);  // we can instantiate modules as usually

DEFINE
  //signals defined in the module can be referred to in the property automata
  a := TRUE;
  b := FALSE;

  writtenA := CPUwrite & valueIn=a & done;
  readA := CPUread & valueOut=a & done;
  is42 := h.reached42;
  ...
  // thus we can use variables `is42', `readA', `writtenA' in property automata below

SYS_AUTOMATON_SPEC // guarantees in the GOAL automata format
  guarantee1.gff; 
  !guarantee2.gff; // `!' signals to negate the automaton

ENV_AUTOMATON_SPEC // assumptions in the GOAL automata format
  assumption1.gff;
  !assumption2.gff;
  ...

\end{lstlisting}
\end{figure}

As in the usual SMV format, it consists of modules and the main module.
In the main module, variables to be controlled by the system are marked with
the comment `{\tt --controllable}' (Mealy-type).
The new sections {\sf ENV\_AUTOMATON\_SPEC} and {\sf SYS\_AUTOMATON\_SPEC} 
contain definitions of the assumptions and guarantees respectively.
Every assumption and guarantee in the corresponding sections is expressed by a
file path to a B\"uchi automaton in the GOAL format~\cite{GOAL}.
A file path can be preceded by `{\sf !}' to indicate that the property
is the negation of the automaton.
These property automata will be converted into SMV modules.

\subsubsection*{Restrictions}
The framework we describe in Section~\ref{sec:scripts} converts a given
specification in the extended SMV format into a deterministic game in the AIGER
circuit format.
AIGER circuits are inherently deterministic and so should be automata
used in sections {\sf SYS\_AUTOMATON\_SPEC} and {\sf ENV\_AUTOMATON\_SPEC}.
We require that:
\begin{itemize}
\item guarantees automata are deterministic (or determinizable),
\item assumptions automata represent safety properties.
\end{itemize}
These conditions are sufficient (but not necessary) for the game to be
deterministic, and are required by the conversion script {\sf spec\_2\_aag.py}
described in Section~\ref{sec:scripts}.

\input{scripts}

%% file: scripts.tex
\section{Conversion into the SYNTCOMP Format}    \label{sec:scripts}

We will convert specifications in the extended SMV format into standard and
extended SYNTCOMP formats. 
Specifications in the standard SYNTCOMP can be given to any synthesis tool from
the SYNTCOMP competition. 
Specification in the extended format can either be converted into the standard 
SYNTCOMP format using {\sf justice\_2\_safety.py}, or can be given to our synthesizer 
{\sf aisy.py} that supports it.

The scripts are available at \url{https://bitbucket.org/art_haali/spec-framework}.

\subsubsection*{Standard and extended SYNTCOMP}
In this section we remind what the standard SYNTCOMP format is and then introduce the extension.

\begin{figure}[h]
\centering
\begin{minipage}{0.35\textwidth}
\includegraphics[width=\textwidth]{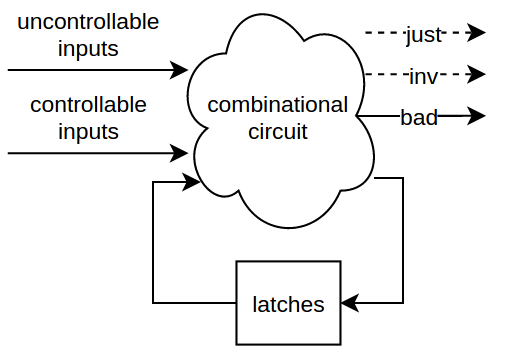}
\captionof{figure}{SYNTCOMP specification}
\label{fig:syntcomp_spec}
\end{minipage}
\hspace{1cm}
\begin{minipage}{0.35\textwidth}
\centering
\includegraphics[width=\textwidth]{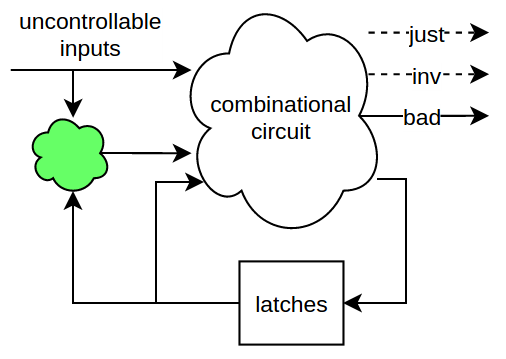}
\captionof{figure}{SYNTCOMP model}
\label{fig:syntcomp_model}
\end{minipage}
\end{figure}

The standard SYNTCOMP is a circuit in the old AIGER format~\cite{AIGEROLD} with
special comments that allow for specifying controllable (by the system) and
uncontrollable (thus controllable by the environment) signals.
Figures \ref{fig:syntcomp_spec} and \ref{fig:syntcomp_model} show the standard
SYNTCOMP format~\cite{SYNTCOMP} (ignore the dotted arrows -- they are part of
the extended format).

The goal is to synthesize the controllable signals (i.e., replace them with
combinational circuits that as inputs use the memory and uncontrollable
signals) such that the output $bad$ never raises.
Thus, the semantics of the standard SYNTCOMP is $\G \neg bad$, which allows for
specifying safety properties.

The natural extension is to allow liveness properties. 
This is what the extended SYNTCOMP format proposes.
It also uses signal $inv$ though it does not add the expressiveness.
These signals are `introduced' using the standard capabilities of the new AIGER
format~\cite{AIGERNEW} (which allows for specifying `bad' signals, `invariant'
signals, and `justice' signals).
The extended SYNTCOMP is shown on the same figure as the standard one if you take 
into account the dotted signals.

The semantics of the extended SYNTCOMP format is 
\begin{equation}
(\neg bad \W \neg inv) \land  (\G inv \impl \GF just) \label{eq:semantics}
\end{equation}

\smallskip\noindent{\bf Note:} the meaning of the signal $just$ is reversed compared to
the new AIGER format~\cite{AIGERNEW}: 
in that case a witness liveness trace satisfies $\G inv \land \GF just$, while
in our case it satisfies $\G inv \land \neg \GF just$.
We reversed the meaning of the signal $just$ to be able to specify properties like $\G(r \impl \F g)$ (``every request is granted'') or $\GF\neg r$ (``request is lowered infinitely often'').
Such properties can be represented by deterministic B\"uchi automata but not by deterministic co-B\"uchi automata. 
And we need specification automata to be deterministic to be able to convert them into inherently deterministic AIGER circuits.

\subsubsection*{Converting specifications into SYNTCOMP}
\begin{figure}[b]
\centering
\includegraphics[width=\textwidth]{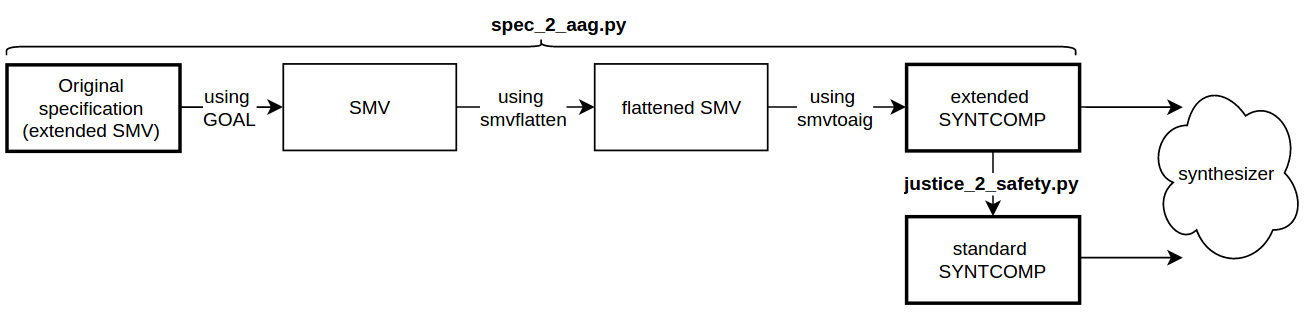}
\caption{Converting specifications from our extended SMV format into the SYNTCOMP format}
\label{fig:conversion}
\end{figure}

Figure \ref{fig:conversion} shows how we convert a given specification into SYNTCOMP format.

The main script is {\sf spec\_2\_aag.py}: 
\li

\-[1.] Given a specification in extended SMV format (Section~\ref{sec:format}),
we first convert all the automata in the GOAL format into SMV modules.
At this step we might need to complement or determinize a given automaton --
this is done using GOAL.
Then we parse the result and convert it into an SMV module.
Such SMV module contains two special signals: $bad$ and $fair$.
In such SMV module, signal $fair$ is risen when we visit an accepting state of
the automaton, and $bad$ is risen when we visit a non-accepting state with 
a self-loop labelled $True$.

\-[2.] The main conversion work -- from the SMV format into the extended
SYNTCOMP format -- is done with scripts {\sf smvflatten} and {\sf svmtoaig} from the
AIGER distribution \cite{AIGER}. 
The result of this step is an AIGER file that may contain invariant and
justice signals, which is not supported by the current SYNTCOMP format.
Thus the current synthesis tools from the competition cannot be used directly.

\-[3.] The file in the extended SYNTCOMP format is converted into the standard
version (with the single output) using {\sf justice\_2\_safety.py}.
The conversion requires input positive integer $k$ and is standard: 
$\GF just$ is replaced with $\G(just \lor \X just \lor ...\X^k just)$, where
$\X^k$ means $k$ repetitions of $\X$.

\il
The result of this conversion is specification in either the standard or extended
SYNTCOMP formats, and can be given to a synthesizer.

\subsubsection*{Converting models into AIGER}
After the synthesizer produces a model, it can be turned into a benchmark in
the standard AIGER format and then be fed to a model checker (e.g., one from
the HWMCC competition):

\begin{figure}[h]
\centering
\includegraphics[width=0.55\textwidth]{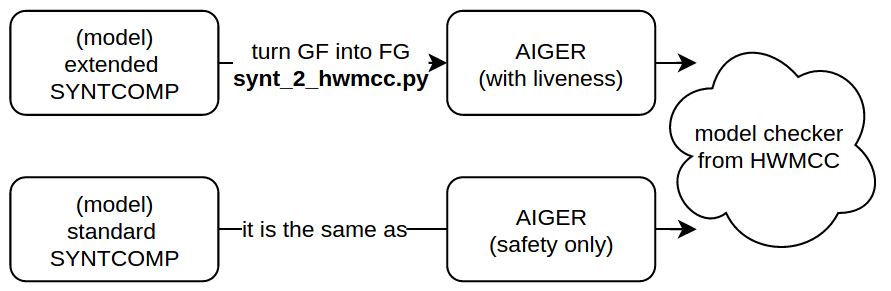}
\end{figure}
If the input synthesis specification is in the standard SYNTCOMP format, then
the model is also in the standard AIGER format and can be fed to a model checker
directly. 
But in the case of the extended SYNTCOMP format we need to translate. 
Recall the semantics of our extended format (Equation~\ref{eq:semantics}): 
in our case a trace violating a liveness property would satisfy $\neg\GF just$,
while the AIGER format has $\GF just'$.
Thus, we convert the model into a model with signal $just'$ such that: 
if there is a trace that satisfies $\GF just'$ then it satisfies $\FG \neg just$.
If denote the new model by $M'$, and the original one by $M$, then: 
$M' \models \pexists \FG just' \impl M \not\models \pforall\GF just$.
The script {\sf synt\_2\_hwmcc.py} does this by introducing a new input $aux$
and attaching the automaton as shown below:
\begin{figure}[h]
\centering
\includegraphics[width=0.6\textwidth]{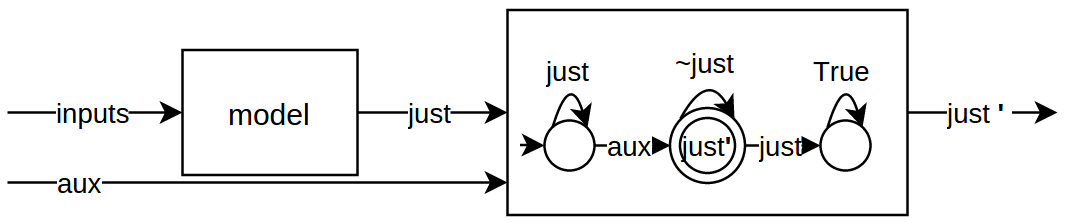}
\end{figure}

%% file: examples.tex
\section{Example: Synthesizing a Huffman Encoder} \label{sec:example}

This section demonstrates the use of the format and the framework. 
We implemented a simple synthesizer that solves B\"uchi games with invariants
and safety objectives given in the extended SYNTCOMP format described in
Section~\ref{sec:format}.
The results of the synthesis are then translated into the HWMCC format using
script {\sf synt\_2\_hwmcc.py}, and then model checked with IIMC~\cite{IIMC}.

We use the Huffman coding \cite{HuffmanCoding} to encode 26 English letters $A...Z$ and the space symbol into bit words of variable length (27 symbols in total). 
Let us assume that a Huffman decoder that decodes a stream of bits into letters is given \footnote{Thus the decoder already has the letter frequencies built in.} ---
the goal is to synthesize an encoder that works with the decoder.

\begin{figure}[t]
\centering
\includegraphics[width=\textwidth]{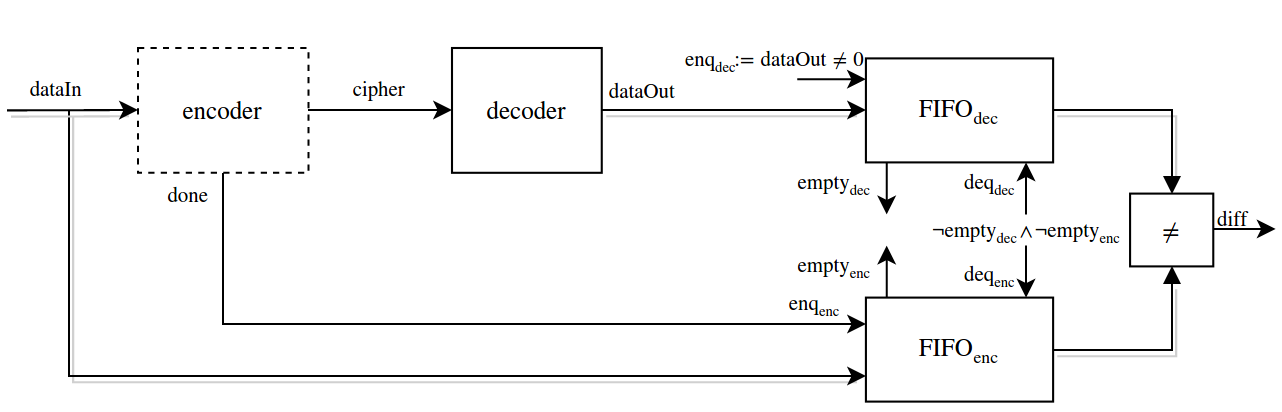}
\caption{The structure of the SMV specificaton for a Huffman encoder}
\label{fig:spec_structure}
\end{figure}

Figure \ref{fig:spec_structure} shows the structure of the SMV specification of the synthesis task.

The dotted module (encoder) is to be synthesized, namely signals $cipher$ and
$done$ (these signals are marked `controllable' in the specification).
The input is $dataIn$ and has five bits of width, which is enough to encode 27
symbol: we use numbers $1..27$ for encoding the symbols.
The outputs of the encoder are boolean signals $cipher$ and $done$; the
intended meaning of $done$ is ``the last bit of the cipher is being sent now''.
The signal $cipher$ is read by the decoder, which decodes the cipher and
outputs it over $dataOut$;
on successful decoding $dataOut$ lasts for one tick, after which it is $0$ again.
The data-signal $dataOut$ is then fed to the FIFO module $FIFO_{dec}$, and $FIFO_{enc}$ takes as input $dataIn$. 
FIFOs values are dequeued whenever they are not empty, and their values are compared.
$FIFO_{enc}$ is enqueued whenever $done$ is high, and $FIFO_{dec}$ -- whenever $dataOut$ encodes a letter.
A FIFO gets blocked if we enqueue and not dequeue, and the FIFO is not empty currently 
(i.e., if $enq \land \neg deq \land empty$ holds).

All modules except dotted module $encoder$ are given: FIFOs we coded manually
(of size 1); the decoder is taken from the distribution of the model checker
VIS~\cite{VIS}.

In words, the specification is:
\li
  \-[A1.] assumption: ``input $dataIn$ is within range $1..27$''

  \-[A2.] assumption: ``$dataIn$ does not change until and including the moment when $done$ is high''

  \-[G1.] $\G (done \impl \X enq_{dec})$
\footnote{Strictly speaking this guarantee is not needed for the correct synthesis of the encoder, but without it the meaning of $done$ may be different from the intended one (``the last cipher bit is being transferred'').}

  \-[G2.] $\G\neg \textit{diff}$, i.e., 
           if FIFOs are not empty, then they contain the same data

  \-[G3.] liveness guarantee: $\GF done$ 
\il
The specification in the SMV format is translated into the SYNTCOMP formats 
(standard safety and extended liveness) as described in Section~\ref{sec:format}.
The semantics is as given in Equation~\ref{eq:semantics} where: $bad$ is the
violation of any of the safety guarantees, $inv$ is the truth of (A1) and (A2)
so far, and $just=done$. 

Given the specification in the extended SYNTCOMP format, the synthesizer 
{\sf aisy.py} synthesized the model in $\approx 2$ minutes; 
the model has $\approx 130$k new AND-gates
\footnote{Recall that we synthesize a memory-less strategy, thus
introduce only new AND-gates and no additional memory.}.
The cipher synthesized is as expected (coincides with that of the Huffman decoder).

If we translate the specification into the $k$-safety variant with $k=10$
(the minimal realizable), then {\sf aisy.py} needs $\approx 4$ minutes 
for the synthesis and the model has $\approx 120$k AND-gates. 
We do not claim that in terms of efficiency the liveness specifications 
are superior to their safety variant -- for this a more thorough research
is needed.  But the translation of liveness into safety requires a value of 
$k$ as input: here we provided it manually, while in the general case its upper 
bound should be restricted and the permitted values should be iterated in some way.

Some final notes on the example.
Initially, FIFOs implementations were non blocking, which permits the
synthesizer to produce a cipher for a letter that is prefixed with ciphers of
other letters (this version of the specification would compare only the last
decoded letter).
Also, with non-blocking FIFOs and without guarantee G2, the synthesizer
produced a cipher that utilized the overflow in the state variable of the
decoder. 
Hence in the general case the synthesized cipher may depend on 
in the implementation of the decoder and will not work with other implementations.

The benchmarks are available as a part of the conversion scripts distribution; 
{\sf aisy.py} is available at \url{https://bitbucket.org/art_haali/aisy}.

%% file: related.tex
\section{Related Work}            \label{sec:related}

\ak{cite AspectLTL -- it is an extension of SMV: \url{http://ysaar.net/data/AspectLTL-aosd11.pdf}}

There are scripts and ways to create specification circuits in the SYNTCOMP format:

The script {\sf ltl2aig} \cite{ltl2aig} takes as input specification in LTL format and signals partition and converts it into a circuit in the standard SYNTCOMP format. 
It does not use tools from the AIGER distribution \cite{AIGER} and supports all the routines natively. 
It also converts liveness properties into safety variants in the standard way.
The limitation is that it does not allow the user to provide partial implementations.

The bundle {\sf ltl2smv}\cite{SMV} - {\sf smvflatten} - {\sf smvtoaig} \cite{smvflatten} can translate SMV files with LTL properties embedded into AIGER format.
The idea is: 
\li
\-[1.] {\sf smvflatten} accepts a given SMV file with modules and variable types like range and enums, and translates it into boolean SMV file, preserving the original {\sf LTLSPEC} section. 

\-[2.] The result is sent to {\sf smvtoaig} that translates {\sf LTLSPEC} section into SMV module using {\sf ltl2smv}, then joins the result, and translates it into AIGER circuit.
\il
I.e, it does what we want but in the context of the model checking. 
For synthesis we also need: 
\li
\- to provide the signals partition (into controllable and uncontrollable) -- a minor issue, and

\- to ensure there are no non-deterministic automata and thus no non-deterministic SMV modules produced at step (2) by {\sf ltl2smv}.
\footnote{This is because we cannot resolve non-determinism by adding the uncontrollable input: the synthesizer is aware of all circuit's signals, thus it may wait for the input to raise and then behave accordingly.
I.e., we need to ensure that a system strategy is independent of the auxiliary signal -- the partial information, which is not supported by the SYNTCOMP format.}
One way to achieve this is to provide a custom implementation of {\sf ltl2smv}.
In hindsight, I think this might be a good way to go.

\il


Finally, in the work in progress paper \cite{Patterns-to-GR1} 
the authors target a similar goal of providing a rich specification language
that benefits from efficient synthesizers.
In that work the authors automatically translate often used LTL patterns 
into the GR(1) fragment of LTL that has an efficient synthesis algorithm~\cite{GR1}.
They do not allow for providing partial implementations.

%% file: conclusions.tex
\section{Conclusions}          \label{sec:conclusions}

In this paper we proposed a format to ease the specification task that allows the user
to provide partial implementations, and we built the conversion scripts from 
the new format into the SYNTCOMP format.  
Both the specification format and the way we convert 
into the existing format are subject to discussion: 
\li
\- Is there a more convenient format of specifications? Is SMV enough or Verilog should be used instead? Should we support GR(1)? Partial information? 

\- Is there a simpler way to convert from the new format into the SYNTCOMP format?
\il

\smallskip
\small\noindent\textbf{Acknowledgements.}
This paper would not be possible without numerous fruitful discussions with 
Robert K\"onighofer, Roderick Bloem, and Georg Hofferek.
Many thanks to reviewers for valuable suggestions.